\documentstyle[pre,aps]{revtex} 
\tighten
\begin{document}
\draft
\twocolumn[\hsize\textwidth\columnwidth\hsize\csname @twocolumnfalse\endcsname
\title{Dynamic transition in deposition with a poisoning species}
\author{F. D. A. Aar\~ao Reis}
\address{
Instituto de F\'\i sica, Universidade Federal Fluminense,\\
Avenida Litor\^anea s/n, 24210-340 Niter\'oi RJ, Brazil}
\date{\today}
\maketitle
\begin{abstract}
In deposition with a poisoning species, we show that the transition to a
blocked or pinned phase may be viewed as an absorbing transition in the
directed percolation ($DP$) class. We consider a ballistic-like deposition
model with an active and an inactive species that represents its basic
features and exhibits a transition from a growing phase to a blocked or pinned
phase, with the deposition rate as the order parameter. In the growing phase,
the interface width shows a crossover from the critical $W\sim t$ behavior to
Kardar-Parisi-Zhang ($KPZ$) scaling, which involves $DP$ and $KPZ$ exponents
in the saturation regime. In the pinned phase, the maximum heights and widths
scale as $H_s\sim W_s\sim  {\left( p-p_c\right) }^{-\nu_{\|}}$. The robustness
of the $DP$ class suggests investigations in real systems.
\end{abstract}

\narrowtext
\vskip2pc]

During some deposition processes, the presence of different chemical species
improves films' properties but may also lead to undesired features, such as the
decrease of growth rates due to erosion processes or the saturation of
dangling bonds at the surface. One important example is the deposition of
$Si$ films doped with $P$ by $CVD$ or $MBE$ in atmospheres with
phosphine~\cite{werner,li,gao}, in which it is observed the decrease of
growth rates when phosphine flux increases. This feature seems to be related to
the saturation of dangling bonds at the surface~\cite{gao}. Similar poisoning
effect appears in diamond $CVD$ in atmospheres with boron and
nitrogen~\cite{edgar}. High fluxes of the poisoning species may cancel out the
growth of the main species, thus showing a transition from a growing phase to
a blocked or pinned phase. Here we will argue that, in the absence of erosion
processes of these two species, it may be viewed as a transition to an
absorbing state in the directed percolation ($DP$)
class~\cite{broadbent,kinzel,marro,hinrichsen}, and we will present a
deposition model that represents the main features of this process.

We will consider a statistical model that represents the essential aspects of
films' growth and may be used to calculate growth rates,
analyse surface roughness scaling and predict a dynamic transition. It is a
ballistic-like deposition model with two species, an active one
($A$) and an inactive one ($B$), with a continuous transition from a
growth phase to a blocked phase. The mapping of this transition onto the $DP$
class shows that the growth velocity is the order parameter of the problem and
that the growth phase corresponds to the active phase of $DP$. Thus their
physical properties are completely different from previous models of surface
growth with pinning or roughening
transitions~\cite{buldyrev,tang,alon,livi,park}. The observed fall of
deposition rates in the growth regime agrees qualitatively with
deposition experiments showing poisoning effects. Thus,
the interpretation of the pinning process as a transition to an absorbing state
and the robustness of the $DP$ class strongly suggest that other transitions to
blocked phases due to poisoning of films' growth are also in the $DP$ class.
Furthermore, we will show that the scaling of quantities such as growth rates,
surface roughness and thicknesses of blocked deposits involve the exponents of
the Kardar-Parisi-Zhang ($KPZ$) theory~\cite{kpz} and $DP$ exponents, and may
eventually be used to compare our theory with experimental data.

In the following we will describe our model, show the results in
one-dimensional substrates while discussing the relation to $DP$, show some
results in two dimensions and present a final discussion.

In our model, particles $A$ and $B$ are released from random positions
above a $d$-dimensional surface of length $L$ with probabilities $1-p$ and $p$,
respectively. The incident particle follows a straight vertical trajectory
towards the surface. Aggregation is allowed only if the incident particle
encounters a particle $A$ at the top of the column of incidence or at the top
of a higher neighboring column. Otherwise the aggregation attempt is
rejected. Fig. 1a illustrates the aggregation rules. A column in which
aggregation is possible will be called an active column. The deposition time
is the number of deposition attempts per substrate column, thus the deposition
rate (number of deposited particles per unit time) is equal to the fraction of
active columns.

It is clear from Fig. 1a that particles $B$
represent impurities that prevent the growth to occur in their neighborhoods.
This model resembles the $AC$ model proposed by other
authors~\cite{wang,elnashar}, but their results are very different from ours
(a morphological transition was suggested in $d=2$~\cite{elnashar}, but it was
not quantitatively studied). Our findings are also completely different from
the two-species $RSOS$ model of Ref. \protect\cite{rsosac}, although
the pure case ($p=0$) also obeyed $KPZ$ scaling.

Now we will present results in $d=1$.

For small values of $p$, the growth process continues indefinitely, such as in
the pure model ($p=0$).However, when $p$ increases, the growth rate $r$
decreases due to the increase in the density of $B$ at the surface, as
shown in Fig. 2a. In Fig. 2b we show $\ln{r}$ versus
$\ln{\left( p_c-p\right)}$ for $p_c=0.20715$, which gives the
best linear fit of the data for $0.19<p<0.206$. Thus we obtain
\begin{equation}
r \sim \epsilon^\beta , \epsilon\equiv p_c-p ,
\label{eq:1}
\end{equation}
with $p_c=0.20715\pm 0.00010$ and $\beta = 0.282\pm 0.012$.

The instantaneous growth rate decays as the density of particles
$A$ at the surface. Focusing on the surface configuration, we
notice that the growth rules of Fig. 1a may be mapped onto a
$d$-dimensional contact process~\cite{harris,marro} ($CP$) in which a top $A$
represents a particle and a top $B$ represents a hole (or empty site), as shown
in Fig. 1b. When the deposition of a $B$ occurs in a column with a top
$A$, it corresponds to the annihilation of a particle in the $CP$. On the
other hand, the deposition of an $A$ in a column with a top $B$ and a
neighboring column with a top $A$ corresponds to offspring production in the
$CP$. Notice that the stability of the absorbing state is
represented by process 4 in Fig. 1b. The probabilities of annihilation and
offspring production in the $CP$ are not trivially related to $p$, since they
also depend on the neighboging heights' distribution.

The equivalence to a $CP$ indicates that the transition is in the $DP$ class,
with the density of top $A$ or the growth rate $r$ as the order parameter.
The above value of the exponent $\beta$ and forward results support this
statement (the best known estimate for $DP$ is $\beta = 0.276486\pm
0.000008$~\cite{jensen}) . 

Here it is relevant to recall that all known statistical models showing
continuous transitions to absorbing states, with positive one-component order
parameters, short-range interactions and no additional symmetries, are in the
$DP$ class~\cite{hinrichsen,grassberger}. This so called {\it robustness} of
the $DP$ class is the reason for us to expect universality in real systems'
transitions with the same blocking mechanisms of our model.

At the critical point in $d=1$ and $L\leq 8192$, we estimated the
deposition rate at very long times, $r_\infty (L)$, and obtained
\begin{equation}
r_c \left( L,t=\infty\right) \sim L^{-\gamma} , 
\label{eq:2}
\end{equation}
with $\gamma = 0.26\pm 0.02$. This result is consistent with the expected $DP$
value $\gamma = \beta/\nu_{\bot}$ (the best known estimate
$\nu_\bot = 1.096854\pm 0.000004$~\cite{jensen} gives $\alpha \approx 0.252$).
We also estimated $r$ for relatively short times in very large substrates
($L=65536$), and obtained
\begin{equation}
r_c \left( L=\infty,t\right) \sim t^{-\eta} ,
\label{eq:3}
\end{equation}
with $\eta = 0.160\pm 0.005$.This estimate also supports the
$DP$ equivalence, which gives $\eta=\beta /\nu_{\|}$, where $\nu_{\|}$ is
the parallel correlation length exponent (best known estimate
$\nu_{\|} = 1.733847\pm 0.000006$~\cite{jensen}).

The interface width, defined as
\begin{equation}
W\left( L,t\right) = {\left[ { \left< { { {1\over{L^d}} \sum_i{ {\left( h_i -
\overline{h}\right) }^2 } } } \right> } \right] }^{1/2} ,
\label{eq:4}
\end{equation}
obeys dynamic scaling involving exponents of $DP$ and $KPZ$ theory
(overbars and angular brackets in Eq. 3 denote spatial and configurational
averages, respectively).
In order to understand its behavior below the critical point, we first show the
results at $p_c$ and very large substrates in Fig. 3. The
interface width $W$ increases as
\begin{equation}
W\sim t , p=p_c ,
\label{eq:5}
\end{equation}
as a consequence of the finite fraction of growing
columns in isolated branches and the increasing fraction of blocked columns,
which give rise to increasingly large heights' differences. 

The evolution of the interface width for $p\lesssim p_c$ is presented in Fig.
4, where we plotted $\ln{W}$ as a function of the
scaling variable $x\equiv t\epsilon^{\nu_{\|}}$, with
$\nu_{\|} = 1.733847$~\cite{jensen}, in substrates with $L=4096$. There is a
transient region for $t<t_{cros}\sim \epsilon^{-\nu_{\|}}$, in which $W$ shows
the rapid increase typical of the critical point (Eq. 5). Notice that
$t_{cros}$ is the characteristic time of correlations in the $DP$ process. At
$t\sim t_{cros}$, $W$ crosses over to a $KPZ$ scaling
\begin{equation}
W\sim t^{\beta_K}
\label{eq:6}
\end{equation}
with $\beta_K =1/3$ in $d=1$. Finite-size effects are responsible for the
reduced declivities in Fig. 4 when compared to the asymptotic forms of Eqs.
(5) and (6) (strong finite-size effects are typical of ballistic deposition
models~\cite{bal}).

For long times, finite-size effects lead to the saturation of the interface
width. The extrapolation of data for several $p$ and $L$, also considering
finite-size effects~\cite{bal}, leads to
\begin{equation}
W_{sat} \sim \epsilon^{-\beta} L^{\alpha_K} , \epsilon\ll 1 , L\gg 1 ,
\label{eq:7}
\end{equation}
with the $KPZ$ exponent $\alpha_K = 1/2$ 
and the $DP$ exponent. $\epsilon^{-\beta}$ is the typical lateral distance
between active columns, but appears in Eq. (4) as a vertical scaling length,
accounting for lateral correlations in the roughness saturation regime. The
divergence of $W_{sat}$ at $p_c$ indicates the failure of $KPZ$
scaling at criticality.

For $p>p_c$, the growth process stops when the whole surface is covered with
$B$, for any length $L$. The heights of the blocked deposits attain limitting
or saturation values with average $H_s$, and the interface widths attain
saturation values $W_s$ ($W_s$ should not be confused with
$W_{sat}$ for $p<p_c$, since the former is a property of infinitely large
static deposits and the latter is related to finite-size effects in growing
deposits). The time for surface blocking is the characteristic time of survival
of particles in the corresponding $CP$, consequently $H_s$ and $W_s$ should
behave as the parallel correlation length in the absorbing phase:
\begin{equation}
H_s \sim W_s \sim {\left( -\epsilon\right) }^{-\nu_\|} .
\label{eq:8}
\end{equation}
Eq. (8) is confirmed in Fig.5, where
we show linear fits of $\ln{H_s}$ and $\ln{W_s}$ versus
$\log{\left( -\epsilon\right)}$, with $p_c =0.20715$ (the same
estimate of the growing phase). From fits with different values of $p_c$ we
obtain $\nu_\| = 1.75\pm 0.05$, which is also consistent with $DP$ within error
bars~\cite{jensen}.

Analogous results were obtained in two-dimensional substrates. In Fig. 6a we
show $\log{r}$ versus $\log{\epsilon}$, with $p_c = 0.4902$, which gives
$\beta = 0.573\pm 0.020$ (Eq. 3). In Fig. 6b we show $\ln{W}$ versus
$\ln{x}$, $x\equiv t\epsilon^{\nu_{\|}}$, for several values of $p$,
considering $\nu_{\|} = 1.295$~\cite{voigt}. Again it shows exponents
consistent with $DP$ and the crossover from $DP$ to $KPZ$ scaling. At $p=p_c$
we obtained Eq. (5) with $\alpha \approx 0.46$, to be compared with the $DP$
value $\alpha \approx 0.451$~\cite{voigt}. The results in $d=2$ are less
accurate due to the limitations in lattice lengths ($L\leq 256$), but are
essential to justify any comparison of our theory with experiments.

The applicability of our model to real growth processes is limitted due to the
ballistic aggregation conditions, the absence of diffusion mechanisms etc.
However, if poisoning effects lead to a transition to a blocked phase and if it
can be interpreted as a transition to an absorbing phase, then the robustness
of the $DP$ class~\cite{grassberger,hinrichsen} suggests this type of
transition. A possible realization is
$Si$ deposition in atmospheres with phosphine ($PH_3$), which shows a decrease
of growth rate with increasing phosphine flux. The saturation of phosphorous
dangling bonds by hydrogen at the surface was suggested as the main blocking
mechanism~\cite{gao}, but to our knowledge no blocking transition was found
yet. Another possible application is diamond CVD in atmospheres with boron, in
which the formation of an amorphous $BCN$ phase blocks the growth of the
diamond phase for boron to carbon ratios above $B/C=0.1$~\cite{edgar}. It seems
to be an absorbing transition similar to our model and, consequently, is a
candidate to the $DP$ class.

Finally, it is important to recall the differences between the transition
found in our model and the pinning transitions by directed percolation of
growing interfaces in disordered
media~\cite{buldyrev,tang,barabasi,hinrichsen}. In that case the
interface is blocked if the impurity concentration exceeds the $DP$ threshold,
then infinite surface growth is found in the absorbing phase of the impurities
system. Consequently, the critical behavior of geometric quantities such
as growth rate and interface width are completely different; for instance,
Eq. (3) is obeyed with $\beta = \nu_\| -\nu_\bot$~\cite{barabasi}. A
very different correspondence to $DP$ is also found in models with competition
between aggregation and desorption that show roughening
transitions~\cite{alon}, in which the film growth regime parallels the
absorbing $DP$ phase.

\acknowledgements

The author thanks Dr. Dante Franceschini for useful suggestions and helpful
discussions. This work was partially supported by CNPq and FAPERJ (Brazilian
agencies).

\begin{figure}
\caption{(a) Examples of deposition attempts in $d=1$, in which only the
configurations of the incident column and of neighboring columns are shown.
Open squares represent particles $A$, filled squares represent particles $B$
and crossed squares represent incident particles ($A$ or $B$). In processes
(1), (2) and (3), aggregation occurs at the positions marked with a filled
circle. In processes (4) and (5) the aggregation attempt is rejected. Notice
that, in processes (3) and (4), lateral aggregation to the right is not
possible because the neighboring $A$ is not at the top of the column. (b) The
equivalent one-dimensional contact process, in which a top $A$ corresponds to a
particle (empty circles) and a top $B$ corresponds to a hole (underlined empty
site). The initial configuration and the possible final configurations (for the
cases of incident $A$ or incident $B$) are shown.}
\label{1}
\end{figure}

\begin{figure}
\caption{(a) Deposition rate $r$ versus probability $p$ of incidence of
particles $B$, in $d=1$. (b) Scaling of $r$ near $p_c=0.20715$.}
\label{2}
\end{figure}

\begin{figure}
\caption{Time evolution of the interface width $W$ at the critical point
$p_c =0.20715$ in $d=1$, for a very large substrate ($L=65536$).}
\label{3}
\end{figure}

\begin{figure}
\caption{$\ln{\left( W\right) }$ versus
$\ln{\left( x\right) }$, with the scaling variable
$x\equiv t\epsilon^{\nu_{\|}}$. From below to above, $p=0.15$, $p=0.17$ and
$p=0.18$ ($L=4096$). The regions of critical DP and KPZ behaviors
are indicated.}
\label{4}
\end{figure}

\begin{figure}
\caption{Saturation height $\ln{\left( H_s\right) }$ (squares) and saturation
width $\ln{\left( W_s\right) }$ (crosses) versus $\ln{-\epsilon}$ in $d=1$,
with $p_c =0.20715$. The solid line is a least squares fit of $H_s$ data,
giving a declivity $\nu_\| \approx 1.75$.}
\label{5}
\end{figure}

\begin{figure}
\caption{(a) Scaling of the deposition rate $r$ near $p_c=0.4902$ in $d=2$,
giving $\beta = 0.573$. (b) $\ln{\left( W\right) }$ versus
$\ln{\left( x\right) }$, with $x\equiv t\epsilon^{\nu\|}$, for $p=0.44$,
$p=0.46$ and $p=0.47$ from below to above ($L=256$).
}
\label{6}
\end{figure}

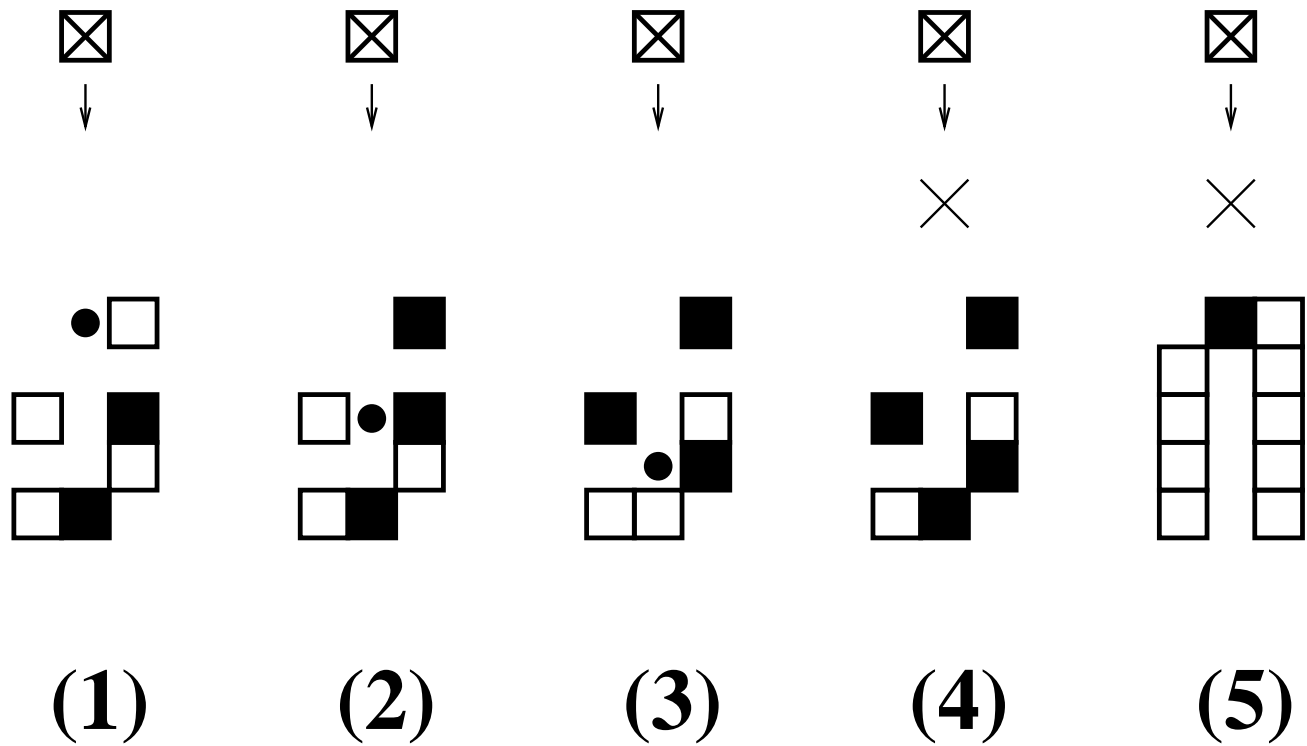
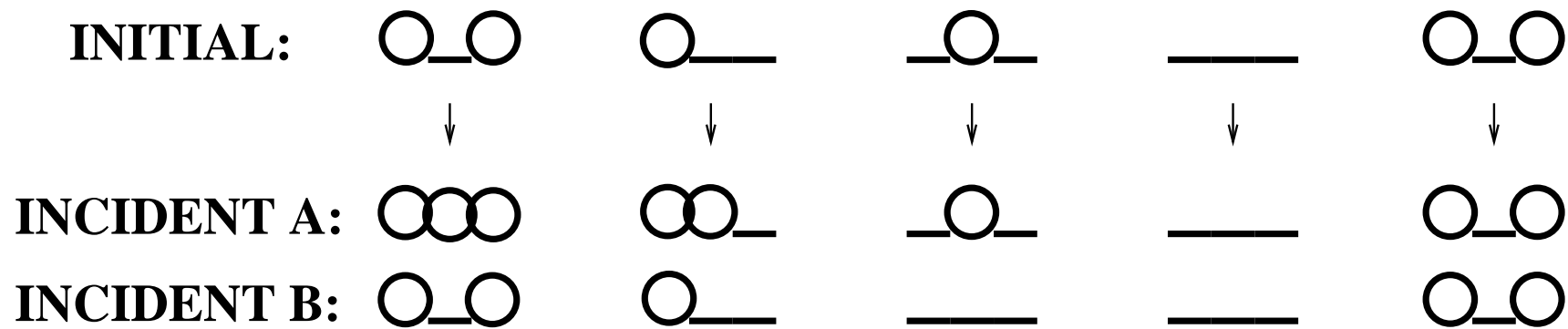
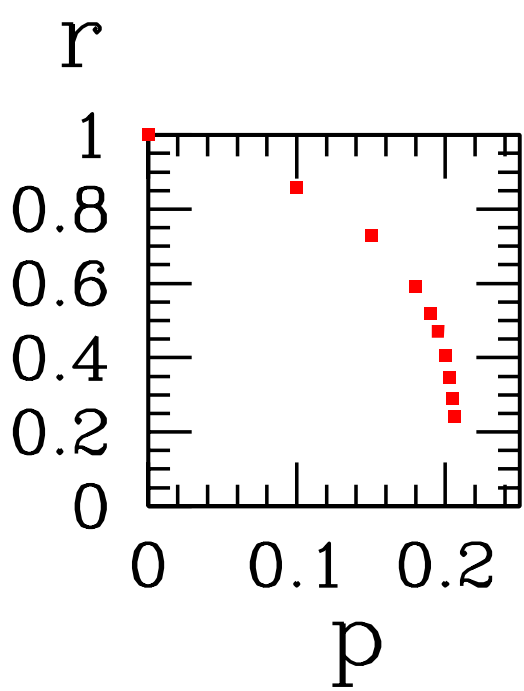
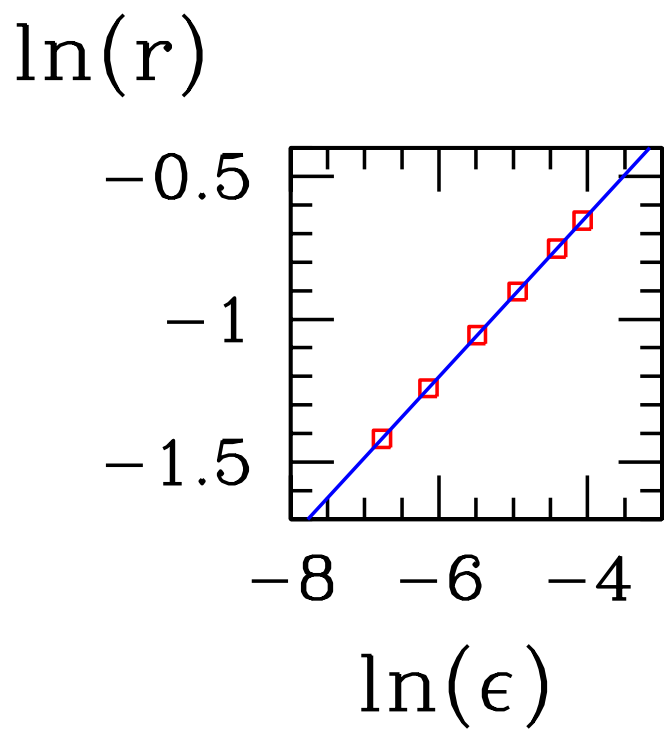
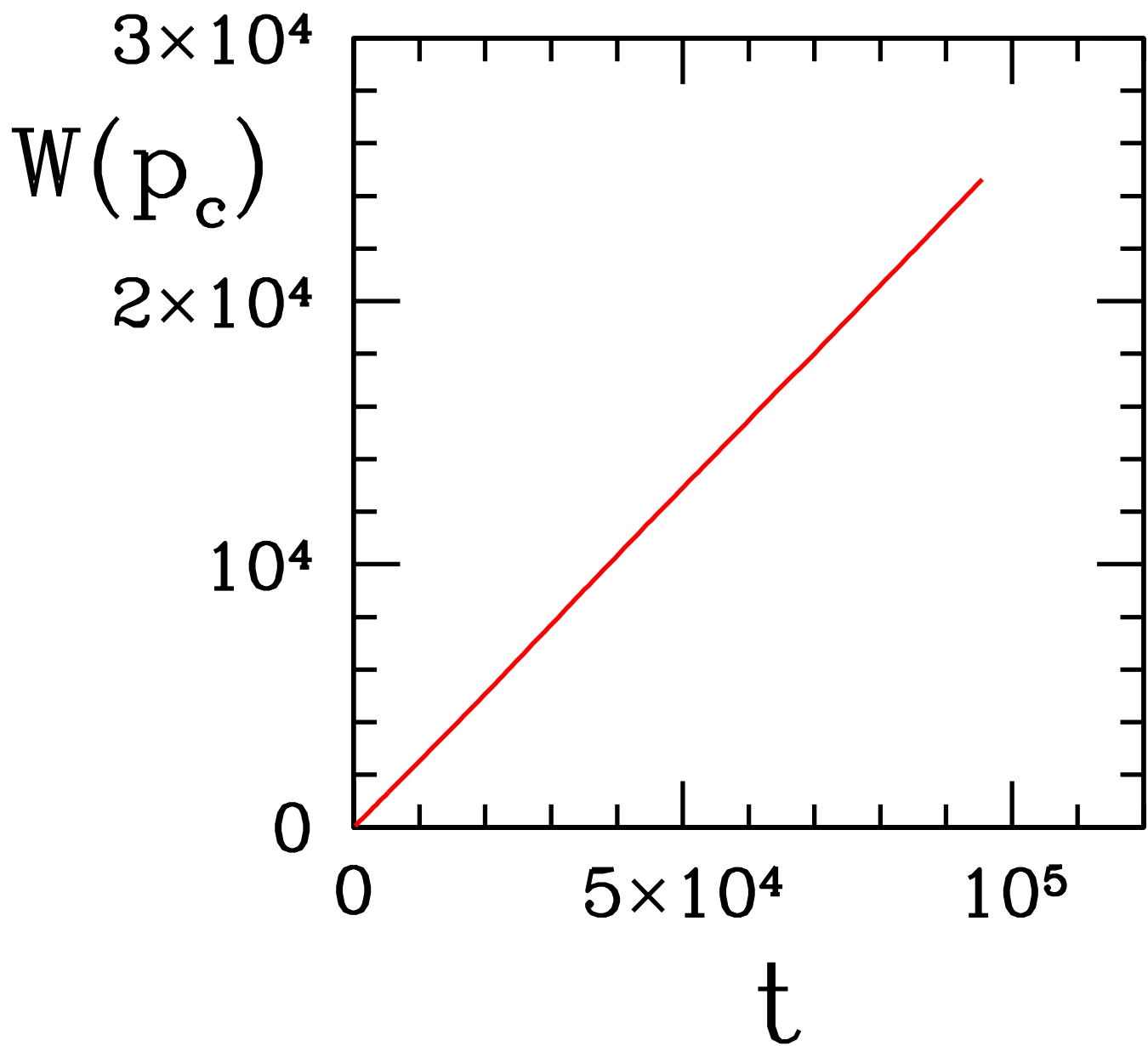
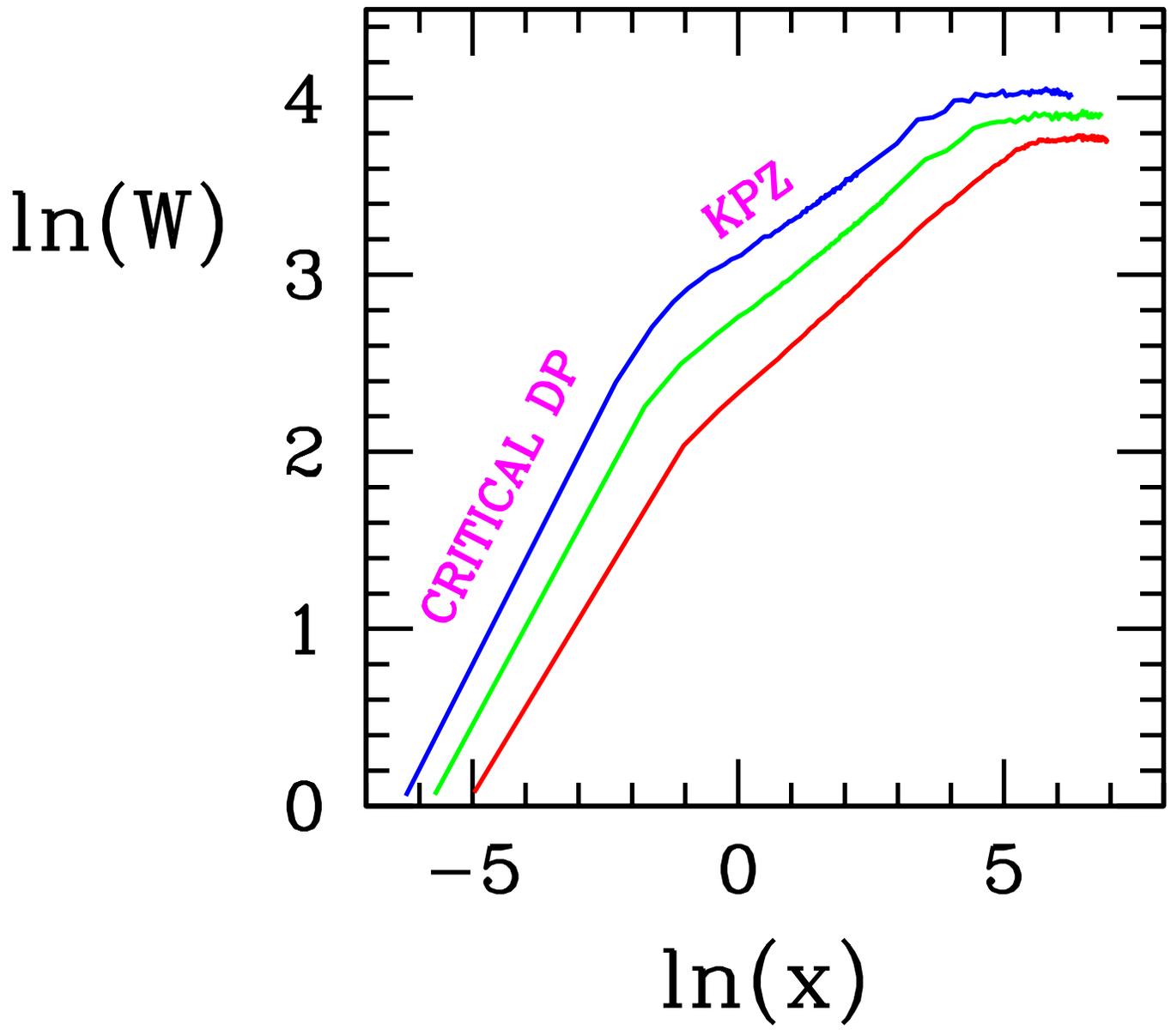
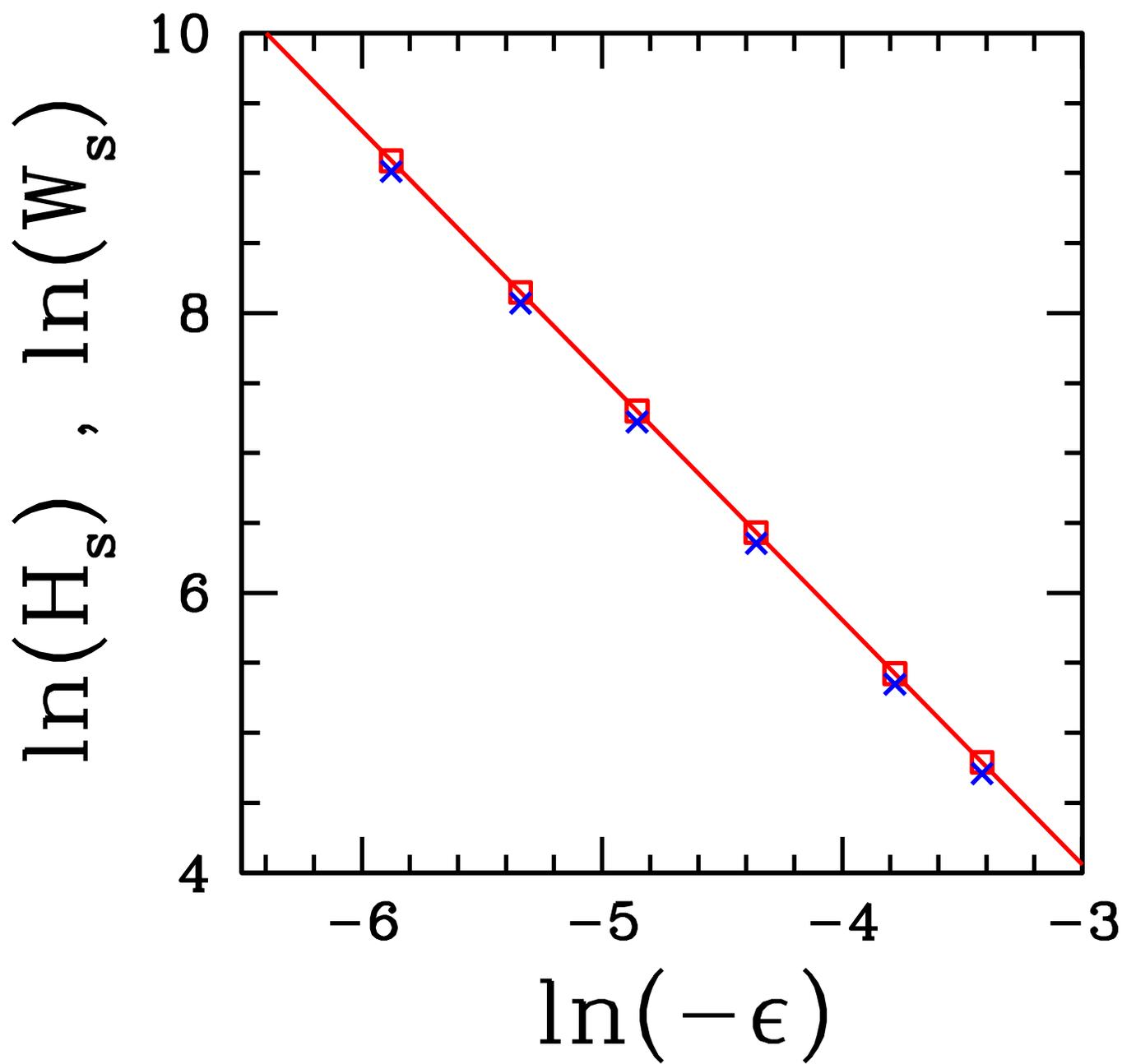
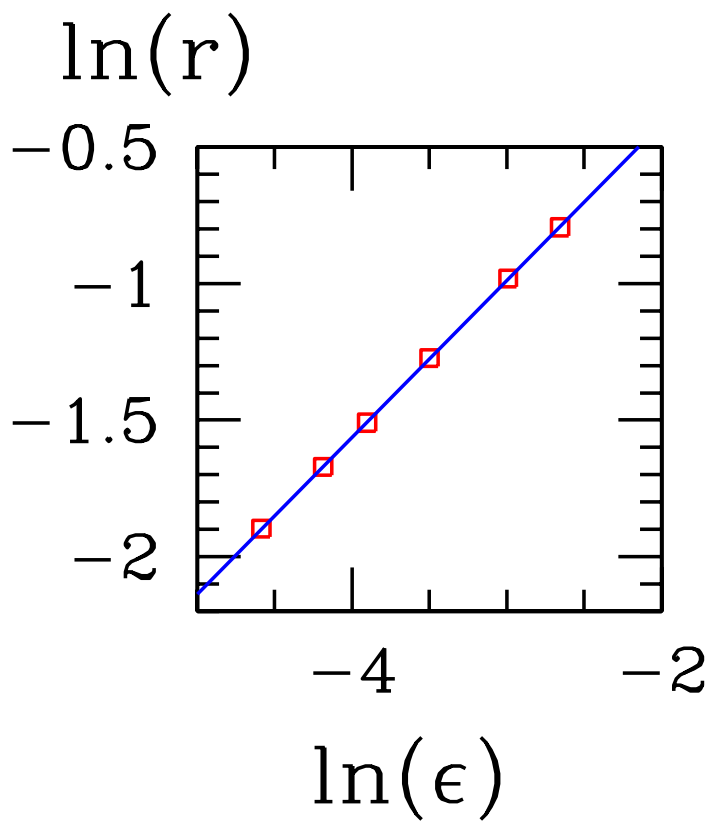
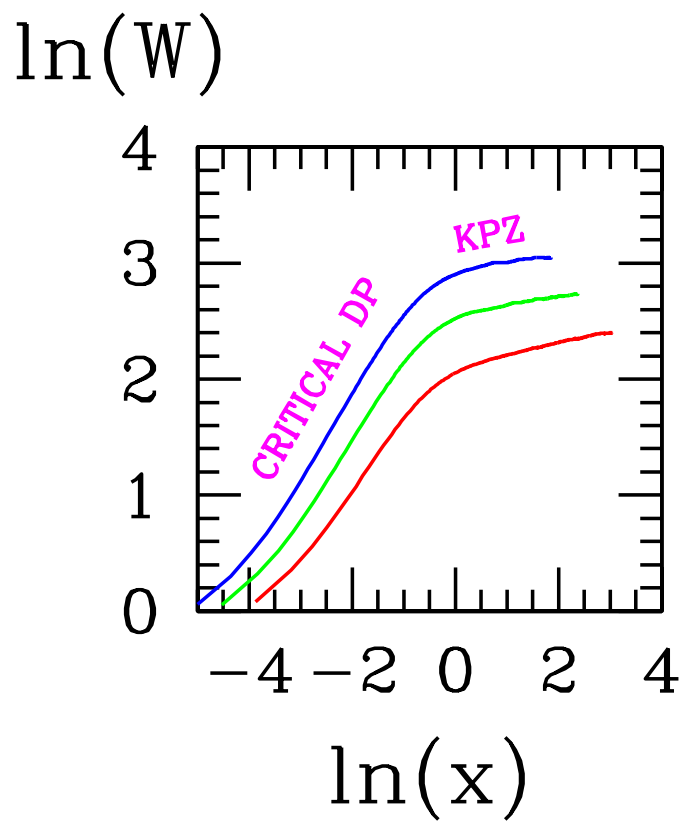
\end{document}